\documentclass[acus]{JAC2003}

\usepackage{graphicx}

\begin{document}
\title{DYNAMICAL EFFECTS DUE TO FRINGE FIELD OF THE MAGNET IN CIRCULAR
ACCELERATORS\thanks{Work partially supported by the Department of
Energy under Contract No. DE-AC02-76F00515.}\vspace{-3mm} }

\author{Y. Cai\thanks{yunhai@slac.stanford.edu}, Y. Nosochkov, SLAC, Menlo
Park, CA 94025, USA}

\maketitle

\begin{abstract}
The leading Lie generators, including the chromatic effects, due
to hard-edge fringe field of single multipole and solenoid are
derived from the vector potentials within a Hamiltonian system.
These nonlinear generators are applied to the interaction region
of PEP-II to analyze the linear errors due to the feed-down from
the off-centered quadrupoles and solenoid.  The nonlinear effects
of tune shifts at large amplitude, the synchro-betatron sidebands
near half integer and their impacts on the dynamic aperture are
studied in the paper.

\end{abstract}

\vspace{-1mm}
\section{INTRODUCTION}
\begin{figure}[htb]
\centering
\includegraphics*[width=80mm]{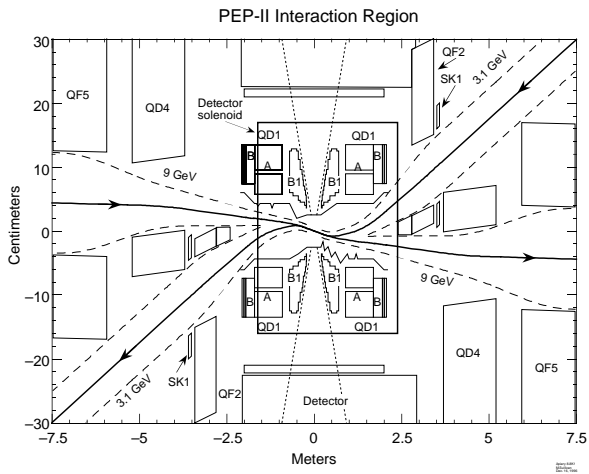}
\vspace{-2mm}
\caption{Top view of the PEP-II magnets and detector solenoid near
the interaction point.} \label{fig:pepii-ip}
\end{figure}

PEP-II is an asymmetric B-factory that consists of two separate rings
with different energies. The electron and positron beams are
brought into head-on collisions at the BABAR detector as shown in
Fig.~\ref{fig:pepii-ip}. In order to separate the beams fast
enough away from the interaction point (IP) to avoid the
deteriorating effect on the luminosity due to adjacent parasitic
collisions, the beams go through many strong magnets inside the
solenoid with very large offsets from their center. These offsets
can be as large as a few centimeters as indicated in
Fig.~\ref{fig:pepii-ip}. The excursion of the design orbit
introduces large uncertainty into the optics near the IP. Most
problematic, the optics changes, as recently seen in the beam-beam
experiments~\cite{witold}, when the local orbit varies.

An accurate optical model requires a map of three-dimensional
magnetic field in the region. That is rather difficult to compute
because of the overlapping fields and complicated geometry. As a
first step, we will use the hard-edge model for the fringe field
to estimate the effects analytically in this paper.

\vspace{-1mm}
\section{VECTOR POTENTIAL}
The magnetic field of a single normal harmonics, including its
fringe field, in the cylindrical coordinate is derived by Bassetti
and Biscari~\cite{bassetti}. To study the single-particle effects
of the fringe field using Hamiltonian system, we need to know its
corresponding vector potential.

\vspace{-1mm}
\subsection{Single normal harmonics, $n>0$}
In the Coulomb gauge, $\nabla\cdot{\vec A} = 0$, the
potential can be expressed as
\vspace{-1mm}\begin{eqnarray}
A_r &=& {\cos n\theta\over2n!}\sum_{p=0}^{\infty}{1\over n+p+1}
G_{n,2p+1}(s)r^{2p+n+1}, \nonumber \\
A_\theta &=& {\sin n\theta\over2n!}\sum_{p=0}^{\infty}{1\over
n+p+1}
G_{n,2p+1}(s)r^{2p+n+1}, \nonumber \\
A_s &=& -{\cos n\theta\over n!}\sum_{p=0}^{\infty}
G_{n,2p}(s)r^{2p+n}, \label{eqn:Amultipole}
\end{eqnarray}
where
\vspace{-1mm}\begin{eqnarray}
G_{n,2p}(s) &=& (-1)^p {n!\over4^p(n+p)!p!}{d^{2p}G_{n,0}(s) \over
ds^{2p}} \nonumber \\
G_{n,2p+1}(s) &=& {dG_{n,2p}(s)\over ds}.
\end{eqnarray}
For a skew multipole, the expression can be obtained by an exchange
between sine and cosine. The Coulomb gauge is chosen because its
potential becomes the conventional multipole expansion.

\vspace{-1mm}
\subsection{Solenoid, $n=0$}
For the solenoid, due to its axial symmetry, it is more convenient
to choose the axial gauge: $A_s = 0$. The vector potential is
given by
\vspace{-1mm}\begin{eqnarray}
A_x &=& -{y\over2}\sum_{p=0}^{\infty} {1\over p+1}
G_{0,2p+1}r^{2p}, \nonumber \\
A_y &=& {x\over2}\sum_{p=0}^{\infty} {1\over p+1}
G_{0,2p+1}r^{2p}. 
\label{eqn:Asolenoid}
\end{eqnarray}
The potential satisfies Maxwell's equation
$\nabla\times\nabla\times{\vec A}=0$. Any truncation of the series
could violate Maxwell's equation. The magnetic field is given by
${\vec B} =\nabla\times{\vec A}$.

\vspace{-1mm}
\section{HARD-EDGE FRINGE}
In the Cartesian coordinate system, the Hamiltonian, using the
distance $s$ as the independent variable for a charged particle
moving in a static magnetic field, is given by~\cite{ruth}
\vspace{-1mm}\begin{eqnarray}
 &&H(x,p_x,y,p_y,\delta,l;s)=-a_s \nonumber \\
 &&-\sqrt{(1+\delta)^2-(p_x-a_x)^2-(p_y-a_y)^2},
\label{eqn:H}
\end{eqnarray}
where $a_{x,y,s}=eA_{x,y,s}/cp_0$ are scaled components of the
vector potential along axis x,y,s, respectively; $p_x, p_x$ are
the transverse canonical momenta scaled by a reference momentum
$p_0$, $\delta=(p-p_0)/p_0$, and $l=vt$ is the path length. We
expand the square root in Eq.~(\ref{eqn:H}) and keep only the
first order of the vector potential,
\vspace{-1mm}\begin{eqnarray}
H = -(1+\delta) + {1\over2(1+\delta)}(p_x^2+p_y^2) \nonumber \\
-[a_s+{1\over 1+\delta}(p_x a_x + p_y a_y)].
\label{eqn:Hamiltonian}
\end{eqnarray}
This Hamiltonian is used to compute the dynamical effects on the
charged particles due to the fringe field.

\vspace{-1mm}
\subsection{Solenoid}
Taking a solenoid with field $B_s$ as an example and follow the method used by
Forest and Milutinovic~\cite{forest}, we choose a hard-edge model
\vspace{-1mm}\begin{equation}
G_{0,1} = B_s\theta(s),
\end{equation}
where $\theta(s)$ is the unit step function. Using this model and
the vector potential in Eq.~(\ref{eqn:Asolenoid}), we have, $A_s =
0$ and
\vspace{-1mm}\begin{eqnarray}
A_x &=& -{y\over2}\{B_s\theta(s)-{B_s\over8}(x^2+y^2)\theta''(s) +
...\} \nonumber \\
A_y &=& {x\over2}\{B_s\theta(s)-{B_s\over8}(x^2+y^2)\theta''(s) +
...\}.
\end{eqnarray}
Substituting these components into the Hamiltonian in
Eq.~(\ref{eqn:Hamiltonian}), we obtain
\vspace{-1mm}\begin{equation}
H = D + V_0\theta(s) + V_2\theta''(s),
\end{equation}
where
\vspace{-1mm}\begin{eqnarray}
D &=& -(1+\delta)+{1\over2(1+\delta)}(p_x^2+p_y^2), \nonumber \\
V_0 &=& {K_s\over 2(1+\delta)}(yp_x-xp_y), \nonumber \\
V_2 &=& {K_s\over16(1+\delta)}(yp_x-xp_y)(x^2+y^2),
\end{eqnarray}
and $K_s = eB_s/cp_0$. After the standard manipulation of map and
integration by parts~\cite{forest}, we obtain the map
\vspace{-1mm}\begin{eqnarray}
&&{\cal M}_s = e^{:[V_2, D]:} = e^{:f_s:}, \nonumber
\end{eqnarray}
where 
\vspace{-1mm}\begin{equation} f_s =
{K_s\over8(1+\delta)^2}[xy(p_x^2-p_y^2) + p_x p_y(y^2-x^2)]
\end{equation}
Here, $:f:g = [f,g]$ and $[~,]$ denotes the Poisson bracket. Note
that ${\cal M}_s$ is invariant under the two-dimensional rotation
around the axis of the solenoid.

\vspace{-1mm}
\subsection{Dipole}
Similarly, for a dipole magnet, we start with
\vspace{-1mm}\begin{equation}
G_{1,0} = B_0\theta(s),
\end{equation}
and set $n=1$ in Eq.~(\ref{eqn:Amultipole}) to obtain the
components of the vector potential of a dipole magnet as follows
\vspace{-1mm}\begin{eqnarray}
A_x &=& {1\over2}(x^2-y^2)\{{1\over2}B_0\theta'(s)+...\}, \nonumber \\
A_y &=& xy\{{1\over2}B_0\theta'(s)+...\}, \nonumber \\
A_s &=& -x\{B_0\theta(s) - {B_0\over8}(x^2+y^2)\theta''(s) +
...\}, \nonumber
\end{eqnarray}
where $B_0$ is the magnetic field of the dipole. The Hamiltonian
is derived by substituting the vector potential into
Eq.~(\ref{eqn:Hamiltonian}). We have
\vspace{-1mm}\begin{equation}
H = D + V_0\theta(s) + V_1\theta'(s) + V_2\theta''(s),
\end{equation}
where
\vspace{-1mm}\begin{eqnarray}
D &=& -(1+\delta)+{1\over2(1+\delta)}(p_x^2+p_y^2), \nonumber \\
V_0 &=& {x\over\rho}, \nonumber \\
V_1 &=& -{1\over
1+\delta}[{p_x\over4\rho}(x^2-y^2)+{p_y\over2\rho}xy], \nonumber \\
V_2 &=& -{x\over8\rho}(x^2+y^2),
\end{eqnarray}
and $1/\rho = eB_0/cp_0$ and $\rho$ is the bending radius of the
dipole magnet. The final map is written as
\vspace{-1mm}\begin{equation}
{\cal M}_d = e^{:-V_1+[V_2, D]:} = e^{:f_d:}, \nonumber
\end{equation}
where,
\vspace{-1mm}\begin{equation}
f_d = {1\over8\rho(1+\delta)}[-x^2 p_x + 2xyp_y-3y^2p_x)].
\end{equation}
As a first-order kick and $\delta=0$, it agrees with the expression
found by Papaphilippou, Wei, and Talman~\cite{talman}.

\begin{figure*}[htb]
\centering
\includegraphics*[width=160mm]{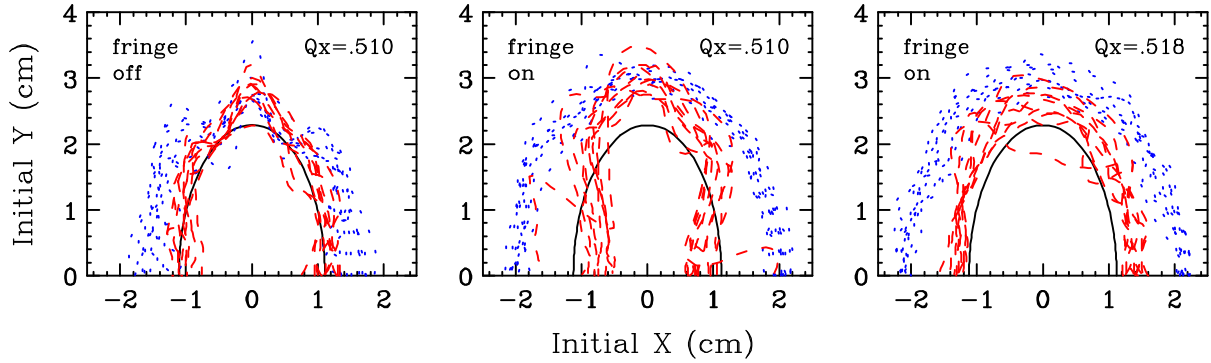}
\vspace{-2mm}
\caption{Evolution of HER dynamic aperture with quadrupole nonlinear fringe 
off and on at $\nu_x=$ .510 and .518.}
\label{fig:aper}
\vspace{-0mm}
\end{figure*}

\vspace{-1mm}
\subsection{Quadrupole}
Similar exercise can be carried out for a quadrupole magnet with
$n=2$. The final map is given by
\vspace{-1mm}\begin{equation}
{\cal M}_q = e^{:f_q:}, \nonumber
\end{equation}
where,
\vspace{-1mm}\begin{equation}
f_q={K_1\over12(1+\delta)}[-(x^3+3xy^3)p_x + (y^3+3yx^2)p_y].
\label{eqn:quadrupole}
\end{equation}
where $K_1 = eG/cp_0$ and $G$ is the gradient. ${\cal M}_q$ agrees 
with the expression first found by Forest and Milutinovic.

Here, the maps are derived for the raising edge of the magnet. For
the falling edge, the maps are obtained by simply switching the
sign of the Lie generator of the map.

If there is a design off-axis orbit: $\Delta x$ and $\Delta p_x$
in the horizontal plane, the non-linear effects of the fringe
field of a quadrupole magnet will feed down to linear optical
errors. By substituting $x$ with $x + \Delta x$ and $p_x$ with
$p_x + \Delta p_x$ into Eq.~(\ref{eqn:quadrupole}), and extracting
the quadratic terms of $x, p_x, y, p_y$, we find that the tune
shifts are given by
\vspace{-1mm}\begin{eqnarray}
\Delta\nu_x &=& {K_1\over 8\pi(1+\delta)}(\Delta x\Delta p_x
\beta_x - \Delta
x^2 \alpha_x), \nonumber \\
\Delta\nu_y &=& {K_1\over 8\pi(1+\delta)}(\Delta x\Delta p_x
\beta_y + \Delta x^2 \alpha_y),
\end{eqnarray}
where $\beta$ and $\alpha$ are the Courant-Snyder parameters at
the position of the edge. For PEP-II, the estimated tune
shifts in the vertical plane relative to the design orbit in the
Low Energy Ring are tabulated in Table~\ref{tab:tuneshift}.

\begin{table}[tb]
\vspace{-2mm}
\begin{center}
\caption{Tune shift from the quadrupole adjacent to the IP.}
\vspace{+0mm}
\begin{tabular}{|l|c|c|c|c|}
\hline \textbf{Name} & s(m) & ${\Delta}x(mm)$ & ${\Delta}p_x$(mrad) 
& ${\Delta}\nu_y$
\\ \hline
QD1L-U    & -2.06 & -30.54 & 46.61 & $8.1\!\times\!\!10^{-3}$ \\
QD1L-D    & -0.90 &  5.01  & 11.38 & $2.3\!\times\!\!10^{-4}$ \\
QD1R-U    &  0.90 & -5.01  & 11.38 & $2.3\!\times\!\!10^{-4}$ \\
QD1L-D    &  2.06 &  30.00 & 46.38 & $8.1\!\times\!\!10^{-3}$ \\ \hline
\end{tabular}
\label{tab:tuneshift}
\end{center}
\vspace{-6mm}
\end{table}

Note that the outside edges of the quadrupole contribute more
because the excursions are larger. These rather large optical
effects are not currently included in our optical model.

\vspace{-1mm}
\section{NONLINEAR EFFECTS}

The nonlinear fringe transformation at quadrupole edges has been recently
implemented in the LEGO code~\cite{lego}.  Based on Eq.~(\ref{eqn:quadrupole}),
the fringe octupole-like field would generate an amplitude dependent tune shift
and excite chromo-geometric octupole resonances.  These effects were observed 
in PEP-II dynamic aperture calculations.  To maximize luminosity, the PEP-II
horizontal tune is moved close to a half-integer.  However, the tune space in
this region is limited by the effects of half-integer resonance and its
synchrotron side bands.  The effect of the quadrupole resonances on PEP-II
dynamic aperture had been observed in earlier tracking
studies~\cite{halfinteger}.  In this case, the resonance condition is
$2\nu_x\!+\!k\nu_s\!=\!n$, where $\nu_s$ is the synchrotron tune.

After including the nonlinear fringe in quadrupoles, the tracking showed a
reduction of dynamic aperture for off-momentum particles.  An example of 
dynamic aperture for PEP-II High Energy Ring (HER) is shown in 
Fig.~\ref{fig:aper}.  In this case, the $90^\circ$ HER upgrade 
lattice~\cite{her90} is used where the IP beta functions and tunes are 
${\beta_x}^*/{\beta_y}^*\!=\!50/1$ cm and 
$\nu_x/\nu_y/\nu_s\!=\!28.51/27.63/0.0405$.  The tracking included synchrotron
oscillations, machine errors and various optics corrections.  In
Fig.~\ref{fig:aper}, the blue dotted lines show on-momentum dynamic aperture 
for 10 random error settings, the dash red lines correspond to relative 
momentum error of $\delta\!=\!8\sigma_\delta$, and the solid ellipse 
shows the $10\sigma$ beam size for reference.  One can see that the fringe 
effect increases the on-momentum aperture but reduces the horizontal 
off-momentum aperture at .510 tune.  The on-momentum improvement is due 
to the fringe compensation of the
amplitude dependent tune shift from sextupoles.  The off-momentum effect was
attributed to the 1st octupole side band of the half-integer resonance excited
by the fringe.  In this case, the octupole resonance condition
$4\nu_x\!+\!l\nu_s\!=\!114$ yields the resonance tune at $\nu_x\!=\!28.5101$.
Moving the tune away from this resonance to 28.518 restored the aperture above
the $10\sigma$ size.

\vspace{-1mm}
\section{CONCLUSION}
We have found a new Lie generator for the hard-edge fringe due to a solenoid 
magnet. It is a fourth-order generator and gives an octupole-like kick to 
charge particles. It can also provide additional x-y couplings through 
an off-centered orbit.


\end{document}